\documentclass[%
 reprint,
 amsmath,amssymb,
 aps,
]{revtex4-1}

\usepackage{graphicx}% Include figure files
\usepackage{dcolumn}% Align table columns on decimal point
\usepackage{bm}% bold math
\begin{document}

%\preprint{APS/123-QED}

\title[Polariton condensation in $S$- and $P$-flatbands in a two-dimensional Lieb lattice]{Polariton condensation in $S$- and $P$-flatbands in a two-dimensional Lieb lattice}% Force line breaks with \\
%\thanks{Footnote to title of article.}

\author{S. Klembt$^{1}$}
\email{sebastian.klembt@physik.uni-wuerzburg.de}
\author{T. H. Harder$^{1}$, O. A. Egorov$^{1}$,  K. Winkler$^{1}$, H. Suchomel$^{1}$,  J. Beierlein$^{1}$,  M. Emmerling$^{1}$, C. Schneider$^{1}$, S. H\"ofling$^{1,2}$}
%\email{sebastian.klembt@physik.uni-wuerzburg.de}

%\author{S. H\"ofling}
\affiliation{
$^{1}$Technische Physik and Wilhelm-Conrad-R\"ontgen Research Center for Complex Material Systems, Universit\"at W\"urzburg, Am Hubland, D-97074 W\"urzburg,
Germany.%\\This line break forced with \textbackslash\textbackslash
}%
\affiliation{
$^{2}$SUPA, School of Physics and Astronomy, University of St Andrews, St Andrews KY16 9SS, United Kingdom%\\This line break forced with \textbackslash\textbackslash
}%

\date{November 1, 2017}% It is always \today, today,

\begin{abstract}
We study the condensation of exciton-polaritons in a two-dimensional Lieb lattice of micropillars. We show selective polariton condensation into the flatbands formed by $S$ and $P_{x,y}$ orbital modes of the micropillars under non-resonant laser excitation.
The real space mode patterns of these condensates are accurately reproduced by the calculation of related Bloch modes of $S$- and $P$-flatbands. Our work emphasizes the potential of exciton-polariton lattices to emulate Hamiltonians of advanced potential landscapes. Furthermore, the obtained results provide a deeper inside into the physics of flatbands known mostly within the tight-binding limit.

%
%Valid PACS numbers may be entered using the \verb+\pacs{#1}+ command.
\end{abstract}

\pacs{Valid PACS appear here}% PACS, the Physics and Astronomy
%Use showkeys class option if keyword
\maketitle

\begin{figure}[b!]
\centering
\includegraphics[width=0.45\textwidth]{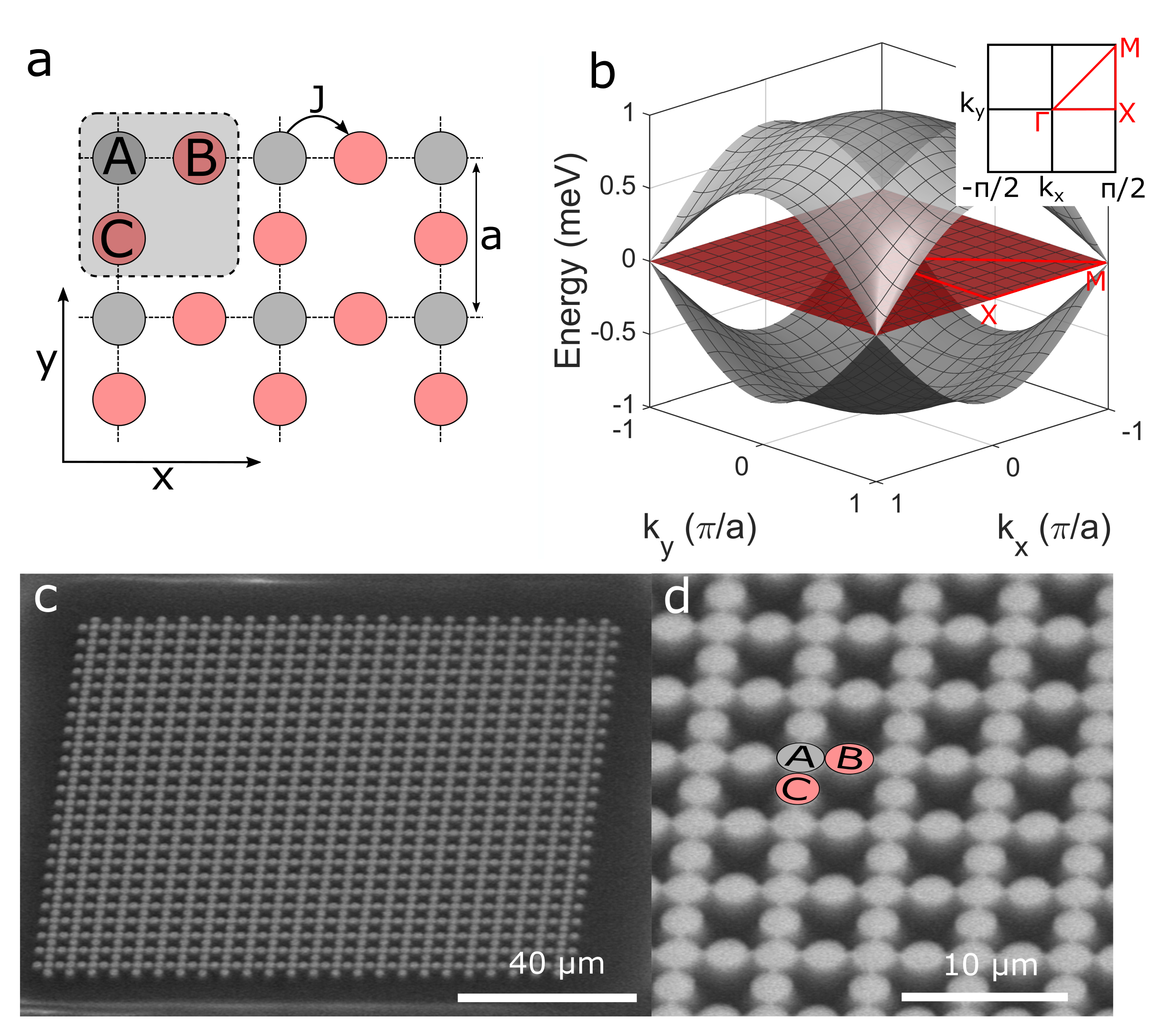}
\caption{(a) Schematic drawing of a Lieb lattice with the three sites A, B, C with a lattice constant $a$ and the unit cell highlighted in a gray box. (b) Tight-binding band structure of a 2D Lieb lattice featuring four Dirac-cone dispersions at the M points as well as a distinct flatband (red). Inset: High-symmetry points $\Gamma$, X, M in the first BZ. Scanning electron microscopy image of a half-etched 2D polariton Lieb lattice (c) and zoom into the structure (d).}
\label{fig:Lieb}
\end{figure}

Dispersionless energy bands or flatbands (FBs) appear in a large variety of condensed matter systems
and are linked to a wide range of topological many-body phenomena  such as graphene edge modes \cite{Wang}, the fractional quantum Hall effect \cite{Wang2, Wang3, Parameswaran, Neupert}
and flat band ferromagnetism \cite{Kusakabe, Tasaki, Tasaki2}.\\
There is a variety of two-dimensional lattices that support flat energy bands \cite{Jacqmin, Masumoto, Gulevich}, with the so-called Lieb lattice being on of the most straightforward examples \cite{Lieb}. Lieb lattices have been studied extensively in recent years and flatband states have been observed in photonic \cite{Guzman, Mukherjee, Vicencio} as well as cold atom systems \cite{Taie}.\\
Creating artificial lattices in order to emulate and simulate complex many-body systems with additional degrees of freedom has attracted considerable scientific interest \cite{Polini, Bloch, Drost}. Exciton-polariton gases in periodic lattice potential landscapes have emerged as a very promising solid state system to emulate many-body physics \cite{Amo, Schneider}. Polaritons are eigenstates resulting of strong coupling between a quantum well exciton and a photonic cavity mode. The excitonic fraction provides a strong non-linearity while the photonic part results in a low effective mass, allowing the formation of driven-dissipative Bose-Einstein condensation \cite{Kasprzak, Balili}. These so-called \textit{quantum fluids of light} \cite{Carusotto} can be placed in an artificial lattice potential landscape using a variety of well developed semiconductor etching techniques \cite{Bayer, Jacqmin, Winkler}, thin metal films \cite{Kim}, surface acoustic waves \cite{Cerda}, or optically imprinted lattices \cite{Ohadi, Berloff}.
\\ In this work we investigate the polariton photoluminescence (PL) emission in a two-dimensional Lieb lattice (Fig. 1(a)). Due to destructive interference of next neighbor tunneling J, flatbands form. Fig.~\ref{fig:Lieb}(b) shows a tight-binding calculation of the first Brillouin zone (BZ) band structure, with the flatband dispersion highlighted in red. High symmetry points of the BZ are found in the inset.

The two-dimensional polaritonic Lieb lattice was fabricated using an electron beam lithography process and a consecutive reactive ion etching step on an AlAs $\lambda$/2-cavity with three stacks of four 13\,nm wide GaAs quantum wells (QWs) placed in the antinode of the electric field, with a $32.5$ ($36$) fold AlAs/Al$_{0.20}$Ga$_{0.80}$As top (bottom) distributed Bragg reflector (DBR) (Fig.~\ref{fig:Lieb}(c,d)). The Rabi splitting of the sample is 9.5\, meV. Only the top DBR is etched in order to create a sufficient potential landscape and not to damage the optically active region of the QWs. The etched micropillars in the Lieb lattices have a diameter of $3.0\,\mu$m and $2.5\, \mu$m at a cavity-exciton detuning of $\delta$=-23.2\,meV  with a normalized next neighbor distance of $v=a/2d=1$, where $a$ is the lattice constant and $d$ the pillar diameter, meaning that in our case the pillars just touch (Fig.~\ref{fig:Lieb}(d)).
The sample was mounted in a liquid Helium flow cryostat with a constant temperature of 6\,K. The non-resonant PL experiments where performed using a pulsed Ti:Sa-Laser with a pulse length of 2\,ps and a repetition rate of 82\,MHz, tuned to a Bragg mode minimum on the high energy side of the stop band at around 1.596\,eV with a spot size of around $25\, \mu$m diameter. The PL emission was collected using a $50\times$ magnification objective with NA=0.42. In order to be able to display modes at a certain energy in real space, mode tomography was performed by shifting the last imaging lens and consecutively taking spectra, allowing for full information in ($x,y$) and energy.\\

First, we calculate the energy-momentum band structure of the Lieb lattices using a full description of the Bloch modes taking into account all relevant system parameters. For this aim, we solve the following eigenvalue problem for the energy $ \hbar \mu ({\bf{k}}_b) $ of the Bloch mode with the Bloch vector $ {\bf{k}}_b =\left\{k_{bx},k_{by}\right\}  $
\begin{equation}\label{eq:eigenvalue}
\hbar \mu \left\{ {\begin{array}{*{20}{c}}	{{p_b}({\bf{r}},{{\bf{k}}_b})}\\
	{{e_b}({\bf{r}},{{\bf{k}}_b})}	\end{array}} \right\} = \hat L({\bf{k}}_b) \left\{{\begin{array}{*{20}{c}}	{{p_b}({\bf{r}},{{\bf{k}}_b})}\\
	{{e_b}({\bf{r}},{{\bf{k}}_b})}	\end{array}} \right\},
\end{equation}
where the functions $ {p_b}({\bf{r}},{{\bf{k}}_b}) $ and $ {e_b}({\bf{r}},{{\bf{k}}_b}) $ describe the amplitude distributions of the photonic and excitonic component of the Bloch modes in real space defined in the plane of the microcavity $ {\bf{r}}=\left\{x,y\right\} $. The main matrix in Eq.~(\ref{eq:eigenvalue}), describing the single-particle coupled states of excitons and photons, is given by the expression
 \begin{equation}\label{}
 \hat L = \left( {\begin{array}{*{20}{c}}
 	{\hbar \omega _C^0 + \hbar V({\bf{r}}) - \frac{{{\hbar ^2}}}{{2{m_C}}}{{\left( {\vec \nabla _ \bot ^{} + i{{\bf{k}}_b}} \right)}^2}\quad \quad \quad \hbar \Omega \quad }\\
 	{\quad \hbar \Omega \quad \quad \quad \quad \quad {\kern 1pt} \quad \hbar \omega _E^0 - \frac{{{\hbar ^2}}}{{2{m_E}}}{{\left( {\vec \nabla _ \bot ^{} + i{{\bf{k}}_b}} \right)}^2}}
 	\end{array}} \right).   \nonumber
 \end{equation}
In the model above, the quantities $ \omega _C^0 $ and $ \omega _E^0 $ represent the energies of bare photons and excitons, respectively. The photon-exciton coupling strength is given by the parameter $ \hbar \Omega $ which defines the Rabi splitting for the micropillars as $ 2\hbar \Omega$ = 9.5\,meV. Here, $ m_C = 32.3 \times 10^{-6}m_e$ is the effective photon mass in the planar region and $ m_e$ is the free electron mass. The effective mass of excitons is $ m_E \approx 10^{5}m_C$. An external photonic potential $V({\bf{r}})$ is defined within the unit cell of the Lieb structure (see Fig.~\ref{fig:Lieb}(a)) compound of micropillars. We assume that the potential is $V({\bf{r}})= 30$ meV outside the micropillars and zero otherwise.            

\begin{figure}[t!]
\centering
\includegraphics[width=0.50\textwidth]{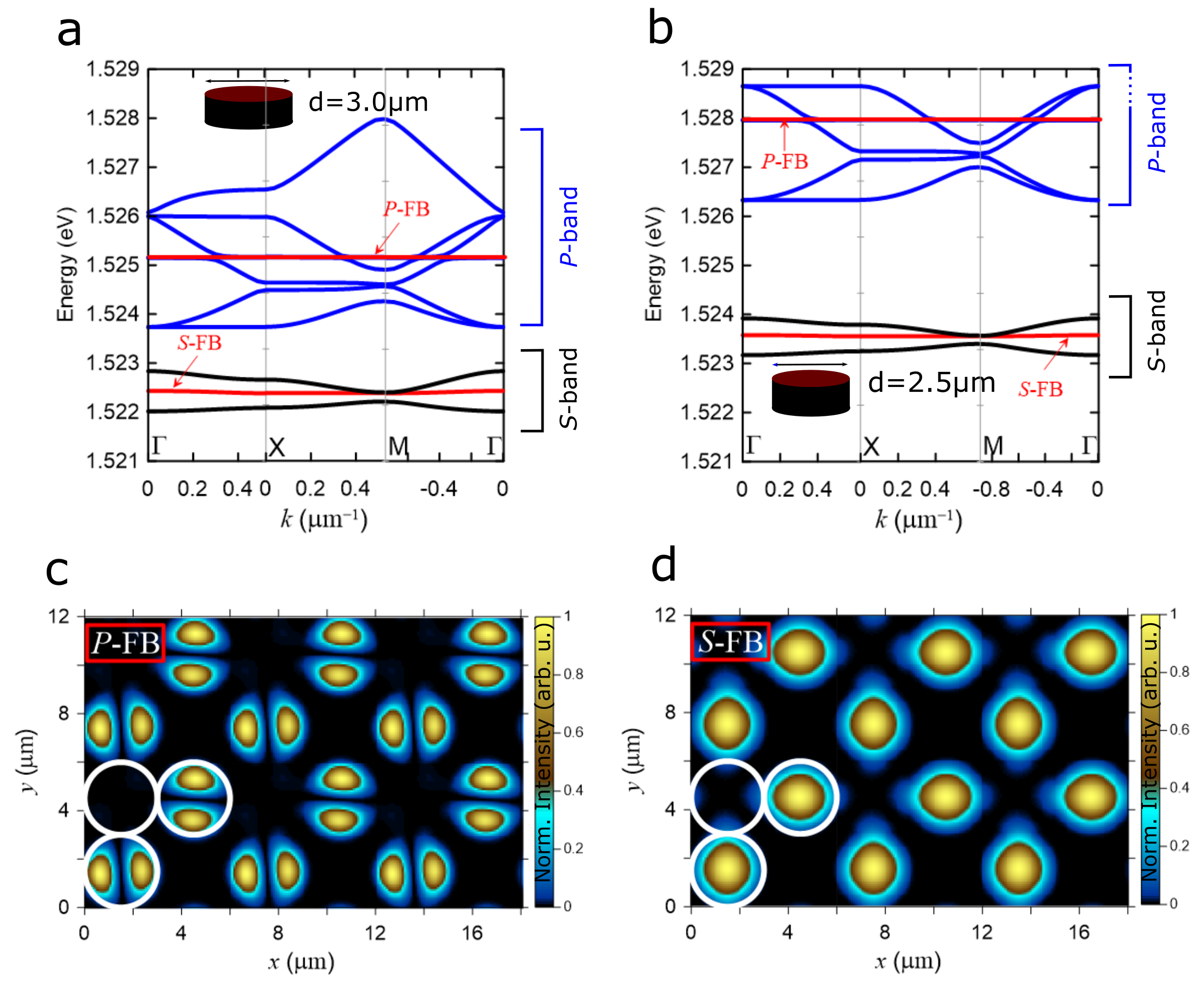}
\caption{Full Bloch-mode calculation of the band structure of a polaritonic two-dimensional Lieb lattice along the high-symmetry points of the first Brillouin zone $\Gamma$-X-M-$\Gamma$ for a pillar diameter d=3.0\,$\mu$m (a) and d=2.5\,$\mu$m (b). (c,d) Real space mode pattern of the $P$- and $S$-flatband, respectively. The white circles indicate the Lieb unit cell.}
\label{fig:Lieb:Theory}
\end{figure}

In Figs.~\ref{fig:Lieb:Theory}(a,b) we plot the bandstucture of the Lieb lattice along the high-symmetry points of the first Brillouin zone $\Gamma$-X-M-$\Gamma$ for a pillar diameter d=3.0\,$\mu$m and d=2.5\,$\mu$m, respectively. 
We mention that the fascinating physics of particle dynamics in the vicinity of flatbands or dirac cones have been mostly studied within the tight-binding approach which is well-justified for the electrons in crystals. However, this description is not necessarily justified in micropillar systems studied here, since the mode profile of the separate pillars can be substantially modified by the neighboring ones. Within a more general approach we found two (at least) dispersive-less bands (flatbands) occurring from a fundamental $ S $ and from the second $ P $ modes of the micropillars (highlighted in red in Figs.~\ref{fig:Lieb:Theory} (a,b) ).

A real space plot of these flatbands (Fig.~\ref{fig:Lieb:Theory} (c,d)) shows the distinct diamond shaped modes where the A-site is dark and light is emitted only from the B- and C-sites. The S-flatband mode is found in exact analogy to localized flatband modes in photonic crystals \cite{Vicencio,Mukherjee}. We point out that Fig.~\ref{fig:Lieb:Theory} (c,d) were calculated for a pillar diameter of d=3.0\,$\mu$m, with the results for d=2.5\, $\mu$m being qualitatively identical with just the length scales changing.
The mode profiles of the $S$-band have slightly elliptical shapes (Fig. 2 d) in spite of the fact that the potential traps are perfect circles.
Evidently, the neighboring sites substantially modify the shapes of the modes, due to direct touching between micro-pillars. Also the shapes of the $P$-band Bloch modes are modified by the neighboring potentials, resulting in the slight tilting of the lobs in respect to the symmetry axes of the Lieb structure (Fig. 2 (c)). These effects definitely go beyond the validity of the tight-binding approximation.  It is worth noting that the $P$-flatband crosses several highly dispersive bands, contrary to what is expected from the tight-binding approach.\\

\begin{figure}[t!]
\centering
\includegraphics[width=0.50\textwidth]{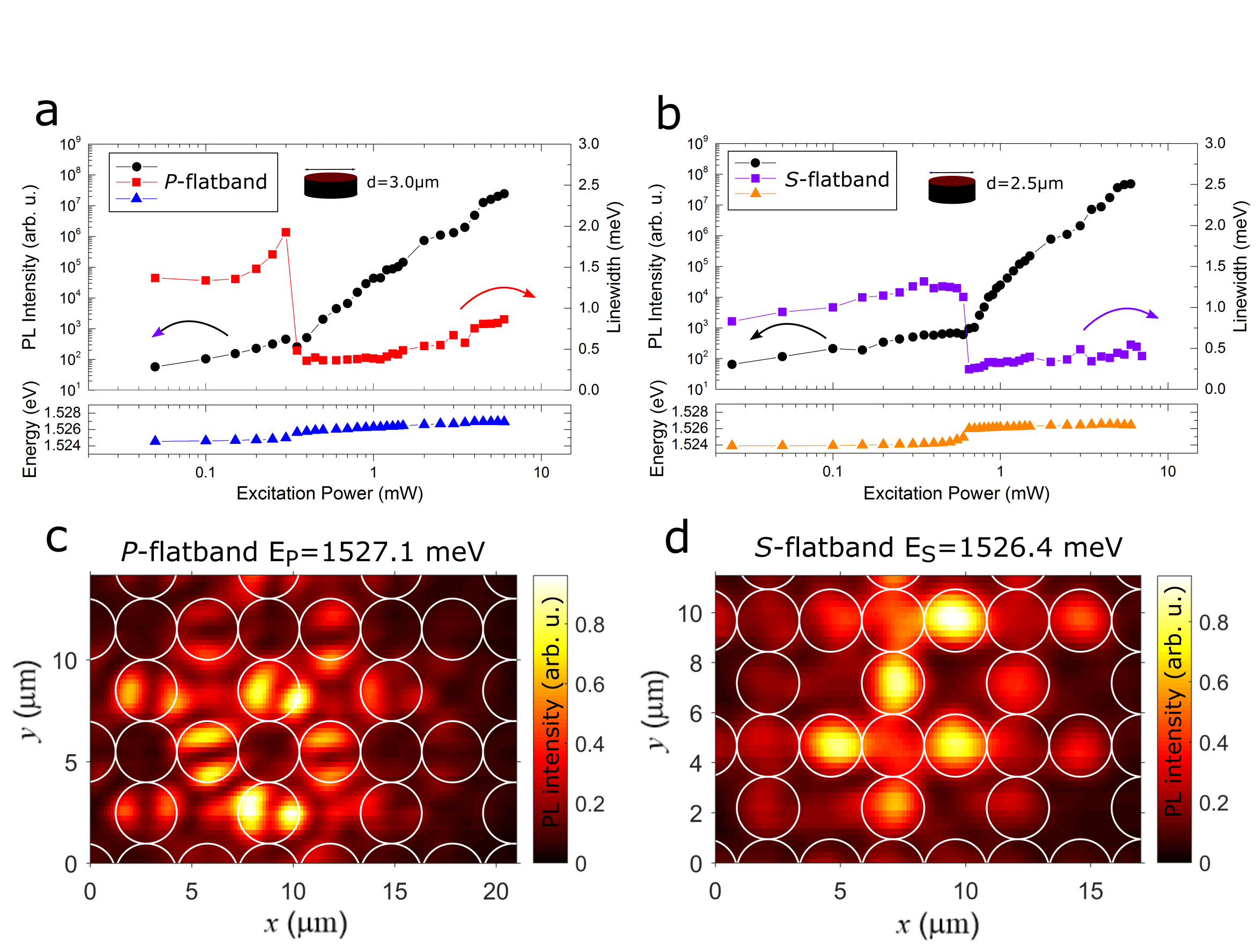}
\caption{(a,b) Peak intensity, emission linewidth and blueshift as a function of the laser excitation power. For a 3.0\,$\mu$m diameter lattice the $P$-flatband emitts dominantly (a), while for a 2.5\,$\mu$m diameter lattice the non-linear emission is dominated by the $S$-flatband. (c) Real space modes of the $P$-flatband at an energy of E$_P$=1527.1\,meV (P$_{Laser}$=3.0\,mW) obtained by mode tomography. (d) Real space modes of the $S$-flatband at an energy of E$_S$=1526.4\,meV (P$_{Laser}$=5.0\,mW). }
\label{fig:Lieb:Experiment}
\end{figure}

In order to study the condensation behavior of our lattice polariton system, we increase the non-resonant pulsed laser excitation with a spot diameter of ~25\,$\mu$m, observing a strong non-linearity of the PL emission intensity accompanied by a sudden decrease of the linewidth (Fig.~\ref{fig:Lieb:Experiment}(a)). For the Lieb lattice with a pillar diameter of 3.0\,$\mu$m and a S-mode-exciton detuning of $\delta$=-22\,meV  the non-linear emission stems dominantly from the P-flatband. 
The real space intensity profile of the condensation mode, plotted in Fig.~\ref{fig:Lieb:Experiment}(c), perfectly matches the calculated intensity pattern of photonic component of the respective Bloch mode in the $P$-flatband (compare with Fig.~\ref{fig:Lieb:Theory}(c)), similar to earlier findings for a one-dimensional Lieb-chain \cite{Baboux}.  Due to the interaction with the incoherent reservoir of hot excitons and carriers, the radiation energy of this mode ($E_P=1527.1$\,meV) is blueshifted about 2.3\,meV with respect to the single-particle band structure.

In order to allow the condesation into the $S$-flatband, we shift to a Lieb lattice with a diameter of 2.5\,$\mu$m at nominally the same cavity-exciton detuning on the sample. Due to the increased confinement potential, the S-mode-exciton detuning is reduced by 2\,meV to $\delta$=-20\,meV. Therefore, the $S$-flatband shifted towards the $P$-flatband energy for the 3.0\,$\mu$m case (see Fig. 2 (a,b)). Here, a dominant non-linear emission from the $S$-flatband at $E_S=1526.4$\,meV is detected (Figs. 3 (b,d)). When shifting to a smaller pillar diameter the increasing confinement in turn increases the excitonic fraction of the polariton. In addition, it has been experimentally shown that spatial confinement enhances phonon-mediated relaxation mechanisms\cite{Paraiso}, which allows for polaritons to relax to the S-flatband for the smaller diameter case. While the overall brightness of the near-field emission varies locally in both flatband polariton condensates, the dominant emission comes from the B- and C-sites in the characteristic diamond shape (compare with Fig. 2(d)). This S-flatband pattern agrees well with the findings of \textit{F. Baboux et al.} for a one-dimensional polariton Lieb chain \cite{Baboux}. \\

\begin{figure}[t!]
\centering
\includegraphics[width=0.40\textwidth]{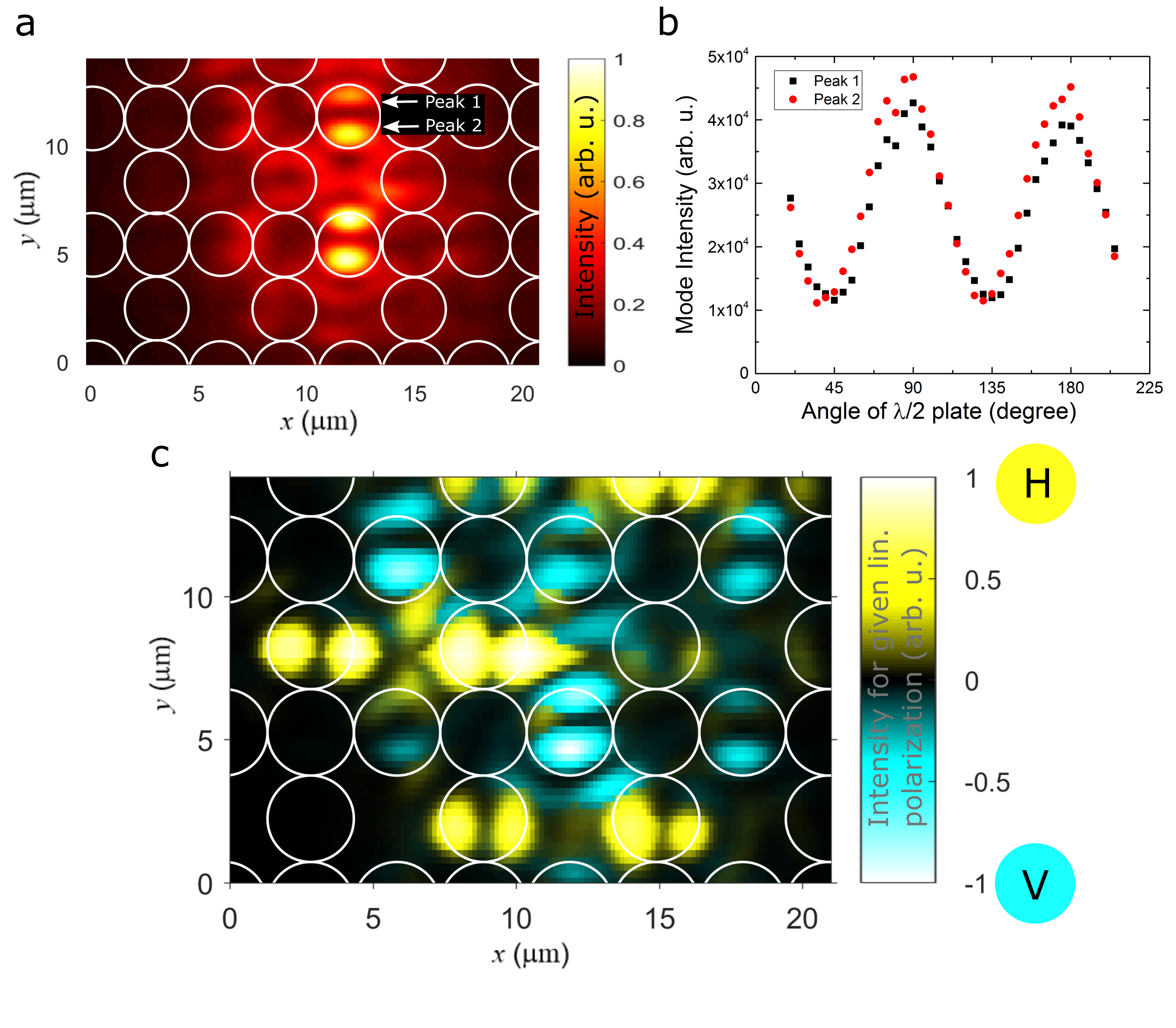}
\caption{(a) Non-energy resolved real space PL of the $P$-flatband using a $\lambda$/2-waveplate and a linear polarizer in the detection path. Primarily the vertically polarized mode is visible. (b) Linear polarization measurement of the $P$-mode (Peak 1 and 2) as a function of $\lambda$/2 rotation angle.(c) Measurement of $P$-flatband real space emission at an energy of E$_P$=1527.1\,meV in horizontal (yellow) and vertical (turquoise) linear polarization.}
\label{fig:Lieb:Polarization}
\end{figure}

In order to study the linear polarization properties of the high angular momentum modes, we introduce a $\lambda$/2-waveplate and a linear polarizer to the detection path in front of the spectrometer. Even without energy filtering the $P$-flatband emission in the d=3.0\,$\mu$m lattice completely dominates the real space emission, shown in Fig. 4(a) for the horizontally polarized part of the spectrum. We then measure the intensity of a P-mode (Peak 1 and 2) as a function of the $\lambda$/2 waveplate angle, showing a clear linear polarization (45$^{\circ}$ periodicity) of the order of 80\%. From this we extract the angle positions for horizontal and vertical polarization degree to perform a mode tomography at the exact energy of the $P$-flatband. As shown in Fig.~\ref{fig:Lieb:Polarization}(c), for the $P$-flatband we find that, while the B-sites are dominantly vertically polarized (turquoise), the C-sites are horizontally polarized. This indicates that the  B-to-A and C-to-A tunnel coupling strongly depends on the polarization state of the polariton. A similar polarization behavior has been found by \textit{C. E. Whittaker et al.} under quasi-resonant excitation in a similar geometry \cite{Whittaker}.\\
\indent In conclusion, we have experimentally demonstrated polariton condensation into the $P$- and $S$-flatband of a two-dimensional Lieb lattice. The real space mode patterns are in excellent agreement with the theoretical data using a full Bloch mode description of the coupled micropillar lattice. We have furthermore demonstrated the possibility to condense into different flatband dispersions selectively, making use of the inherent detuning dependent condensation properties of polaritons. These results underline the potential of exciton-polariton lattices as a non-linear photonic simulator in general and for the emulation of flatband system in particular.

\subsubsection*{Supplementary Material}
\noindent In supplementary figure S1 PL and white light reflectivity data is presented to show the strong light-matter coupling in the microcavity. Supplementary figure S2 shows the $P$-flatband ($S$-flatband) dispersions in the d=3.0\,$\mu$m (d=2.5\,$\mu$m) lattices. 

\subsubsection*{Acknowledgments}
\noindent S.K.  acknowledges  the  European  Commission  for  the H2020 Marie Sk\l odowska-Curie Actions (MSCA) fellowship (Topopolis).
The W\"urzburg group acknowledges the financial  support  by  the  state  of  Bavaria  and  the  Deutsche Forschungsgemeinschaft (DFG) within the project Schn1376-3.1. We would like to thank I. G. Savenko and M. Sun for discussions.

%\newpage

\end{document}